\documentstyle[12pt]{article}
\begin{document}
\setlength{\oddsidemargin}{-1in}
\addtolength{\oddsidemargin}{35mm}
\setlength{\textwidth}{6in}
\setlength{\topmargin}{-1in}
\setlength{\headheight}{35mm}
\setlength{\headsep}{0mm}
\setlength{\textheight}{232mm}
\setlength{\textheight}{8.5in}
\pagestyle{empty}
\title{\bf Leptogenesis and Low Energy CP Violation
\footnote{Talk given by MNR at the RTN meeting :
``Across the Present Energy Frontier : Probing the
Origin of Mass'', Corfu, Greece, 10 September- 13 September 2001}}

\author{G. C. Branco$^a$ \footnote{e-mail:
gbranco@alfa.ist.utl.pt}\ ,
T. Morozumi$^b$  \footnote{e-mail:
morozumi@theo.phys.sci.hiroshima-u.ac.jp}\ ,
B. M. Nobre$^a$  \footnote{e-mail:
bnobre@cfif.ist.utl.pt}\
and M. N. Rebelo$^a$  \footnote{e-mail:
rebelo@alfa.ist.utl.pt}\ }
\maketitle
\begin{center}
$^a$Departamento de F\'\i sica
Instituto Superior T\'{e}cnico, \\ Av. Rovisco Pais P-1049-001,
Lisboa, Portugal\\
$^b$Graduate School of Science, Hiroshima University
1-3-1 Kagamiyama, Higashi Hiroshima - 739-8526, Japan\\
\end{center}

\begin{abstract}
We discuss
the possibility of relating the size and sign of the
observed baryon asymmetry of the universe to CP
violation observable at low energies, in a framework
where the observed baryon asymmetry is produced by leptogenesis
through out of equilibrium decay of heavy Majorana neutrinos.
We have shown that although in general such a connection cannot
be established, there are specific frameworks where a
link does exist. Furthermore, we identify the CP
violating phases relevant to leptogenesis and those relevant
for low energy CP violation and build weak basis invariant
conditions for CP conservation.
\end{abstract}
\section{\bf Introduction}

In the Standard Model (SM) neutrinos are massless and there
is no CP violation in the leptonic sector. However at present
there is strong evidence for neutino oscillations reported by
the SuperKamiokande experiment \cite{SK} and recently confirmed
by the results of the Sudbury Neutrino Observatory (SNO)
\cite{SNO}, both pointing towards nonzero neutrino masses.
The most straightforward way of extending the SM in order to
incorporate neutrino masses is to add one right-handed
neutrino field per generation, singlet under the
$SU(3)_{c} \times SU(2) \times U(1)$ gauge symmetry, in
analogy with the quark sector. In Grand Unified Theories (GUTs),
such as $SO(10)$ GUTs \cite{so}, these right-handed neutrino states
appear in irreducible representations, together with quarks
and leptons. Although this might look like a trivial extension
of the SM it is far reaching in its consequences, giving rise to
entirely new phenomena in the leptonic sector, due to the fact
that neutrinos are neutral particles. In fact, if lepton number
conservation is not imposed, a Majorana mass term for the
neutral right-handed gauge singlets must be included in the
Lagrangean, together with the usual Dirac mass term, leading to
the seesaw mechanism \cite{seesaw} which accounts, in an elegant
way, for the smallness of neutrino masses.
Furthermore mixing and CP violation in the leptonic sector
naturally arise once right-handed neutrino singlets are
included \footnote{It was shown long ago
that mixing and CP violation in the leptonic sector
can also occur with strictly massless neutrinos in a model
where, in addition to right-handed neutrinos, an equal
number of gauge singlet leptons are included
\cite{ago}}.
CP violation in the leptonic sector can have profound
cosmological implications leading to the generation of the
observed Baryon number Asymmetry of the Universe (BAU) via
Leptogenesis. In this framework, the starting point is a CP
asymmetry generated through out-of-equilibrium L-violating
decays of the heavy Majorana neutrinos \cite{Fukugita:1986hr}
leading to a lepton asymmetry L$\neq$0 while B=0 is still
maintained. Subsequently, sphaleron processes \cite{Kuzmin:1985mm},
which are (B+L) violating and (B-L) conserving restore
(B+L)=0 thus creating a nonvanishing B. This mechanism
is, at present, one of the most appealing scenarios
for Baryogenesis. Leptogenesis
has been studied in detail by several groups \cite{buch}
and it has been shown that the observed BAU of
$n_B/s \sim 10^{-10}$ can be obtained in the above
scenario without any fine-tuning of parameters.
In our work \cite{Branco:2001pq} we address the question of
whether it is possible
to establish a connection between CP breaking necessary to
generate leptogenesis and CP violation at low energies.
More specifically, assuming that
baryogenesis is achieved through leptogenesis, can one infer the
strength of CP violation at low energies from the size
and sign of the observed BAU?
We start by studying the various sources of CP violation
in the minimal seesaw model (i.e.,no left-handed Majorana mass
terms) identifying both the CP violating
phases and the weak-basis (WB) invariants which are associated
to leptogenesis and those relevant for CP violation at low
energies. We proceed by showing that this connection is not possible in
general, but we present special scenarios where the connection
can be established. Several authors have addressed this question
under different assumptions \cite{link}.

\section{\bf Framework}
Let us consider a minimal extension of the SM which consists
of adding to the standard spectrum one right-handed neutrino per
generation. After spontaneous gauge symmetry breaking the
leptonic mass terms can be written as
\begin{eqnarray}
{\cal L}_m  &=& -[\overline{{\nu}_{L}^0} m_D \nu_{R}^0 +
\frac{1}{2} \nu_{R}^{0T} C M_R \nu_{R}^0+
\overline{l_L^0} m_l l_R^0] + h. c. = \nonumber \\
&=& - [\frac{1}{2}  n_{L}^{T} C {\cal M}^* n_L +
\overline{l_L^0} m_l l_R^0 ] + h. c.
\label{lm}
\end{eqnarray}
where $m_D$, $M_R$ and $m_l$ denote the neutrino Dirac mass matrix,
the right-handed neutrino Majorana mass matrix and the charged
lepton mass matrix, respectively, and
$n_L = ({\nu}_{L}^0, {(\nu_R^0)}^c)$. Obviously in this minimal
extension of the SM a term of the form
$\frac{1}{2} \nu_{L}^{0T} C m_L \nu_{L}^0$
does not appear in the Lagrangean and the matrix ${\cal M}$
is given by:
\begin{equation}
{\cal M}= \left(\begin{array}{cc}
0 & m \\
m^T & M \end{array}\right) \label{calm}
\end{equation}
with a zero entry on the (11) block.
For notation simplicity, we have dropped the subscript in $m_D$
and $M_R$. It has been shown that in the most
general case, when ${\cal M}$ includes $m_L$,
the number of CP violating phases in ${\cal M}$ is given by
\cite{Branco:gr}:
\begin{equation}
N_{CP} = n n^{\prime} +\frac{n(n-1)}{2}
\end{equation}
where $n$ is the number of $\nu _L$ fields and $n^\prime$ the
number of $\nu _R$ fields. In our case $n = n^\prime$. Without
$m_L$ the number of independent CP violating phases was
computed in Ref.\cite{Endoh:2000hc} to be:
\begin{equation}
n_{CP} = n^{2} - n
\end{equation}
For defineteness, we shall consider the case of three generations
(three light neutrinos). In this case the number of
physical parameters contained in ${\cal L}_m $ is a total of
fifteen real parameters and six CP violating phases as can be
easily seen by going to the weak-basis (WB) where $m_l$ and $M$
are chosen to be diagonal and real matrices. Of course there
is no loss of generality in the choice of a weak-basis. In this
WB, there will be six real parameters in $m_l$ and $M$, on the other
hand $m$ is a general three-by-three matrix and can be written
as the product of a unitary times a Hermitian matrix:
\begin{equation}
m=U H = P_{\xi} {\hat U_{\rho}} P_{\tau}
{P_{\beta}^\dagger}\ {\hat {H_\sigma }}\ P_{\beta}
\label{muh}
\end{equation}
with $P_{\xi}={\rm diag.}\left(\exp(i\xi_1),\exp(i\xi_2),\exp(i\xi_3)
\right)$, $P_\tau ={\rm diag.}(1, \exp(i\tau_1), \exp(i\tau_2))$ and
$P_\beta ={\rm diag.}(1, \exp(i\beta_1), \exp(i\beta_2))$. In the
second equality we have factored out of U and H
as many phases as possible
leaving ${\hat U_{\rho}}$ and ${\hat {H_\sigma }}$ with only
one phase each. Since $ P_{\xi}$ can be rotated away by a WB
transformation corresponding to a simultaneous phase redefinition
of the left-handed charged lepton fields, and the
left-handed neutrino fields, the matrix $m$ is left with six independent
phases and nine real parameters. We shall denote the six independent
phases as $\rho$ in  ${\hat U_{\rho}}$, $\alpha_1$, $\alpha_2$
in the product $(P_{\tau} {P_{\beta}^\dagger})$, $\sigma$ in
${\hat {H_\sigma }}$ and $\beta_1$, $\beta_2$ in $P_{\beta}$.

\section{\bf CP violating phases relevant for leptogenesis}
Leptogenesis gives rise to the BAU through  out-of-equilibrium
decay of heavy Majorana neutrinos in the symmetric phase (i.e.,
before spontaneous gauge symmetry breakdown). The computation
of the lepton number asymmetry, in this extension of the SM,
resulting from the decay of a heavy Majorana neutrino $N^j$
into charged leptons $l_i^\pm$ ($i$= e, $\mu$ , $\tau$) can be
done both in the symmetric phase \cite{sym} and in the broken
phase \cite{Endoh:2000hc}, \cite{Branco:2001pq}.
We define the lepton family number asymmetry as
$\Delta {A^j}_i={N^j}_i-{{\overline{N}}^j}_i$.
The lepton number asymmetry from j th heavy Majorana particle
is then given by:
\begin{equation}
A^j = \frac{\sum_i \Delta {A^j}_i}{\sum_i \left({N^j}_i +
\overline{N^j}_i \right)}
\label{jad}
\end{equation}
with the sum in $i$ running over the three flavours
$i$ = e $\nu$ $\tau$.
In this framework the calculations lead to:
\begin{eqnarray}
A^j &=& \frac{g^2}{{M_W}^2} \sum_{k \ne j} \left[
{\rm Im} \left((m^\dagger m)_{jk} (m^\dagger m)_{jk} \right)
\frac{1}{16 \pi} \left(I(x_k)+ \frac{\sqrt{x_k}}{1-x_k} \right)
\right]
\frac{1}{(m^\dagger m)_{jj}}, \nonumber \\
&=& \sum_{k \ne j} \left[
{\rm Im} \left(({y_D}^\dagger y_D)_{jk} ({y_D}^\dagger y_D)_{jk}
\right)
\frac{1}{8 \pi} \left(I(x_k)+ \frac{\sqrt{x_k}}{1-x_k} \right)
\right]
\frac{1}{({y_D}^\dagger y_D)_{jj}}, \label{rmy}
\end{eqnarray}
The second equality results from the substitution
$m_{ij}=y_{D ij} \frac{v}{\sqrt{2}}$, with $y_{D ij}$ denoting
the coefficients of the neutrino Yukawa couplings and $v$ the
Higgs vacuum expectation value. The variable $x_k$
is defined as  $x_k=\frac{{M_k}^2}{{M_j}^2}$ and
$ I(x_k)=\sqrt{x_k} \left(1+(1+x_k) \log(\frac{x_k}{1+x_k}) \right)$.
From Eq.(\ref{rmy}) it can be seen that the lepton number
asymmetry is only sensitive to the CP violating phases
appearing in $m^\dagger m$. With the choice of phases
of Eq.(\ref{muh}) leptogenesis is only sensitive to $\sigma$,
$\beta _1$ and $\beta _2$.

\section{\bf Weak-basis invariants and CP violation}
The most general CP transformation of the leptonic fermion fields,
still in a WB, which leaves the gauge interaction invariant is of
the form
\begin{eqnarray}
{\rm CP} l_L ({\rm CP})^{\dagger}&=&U^\prime
\gamma^0 {\rm C} \overline{l_L}^T
\quad
{\rm CP} l_R({\rm CP})^{\dagger}=V^\prime
\gamma^0 {\rm C} \overline{l_R}^T
\nonumber \\
{\rm CP} \nu_L ({\rm CP})^{\dagger}&=&U^\prime
\gamma^0 {\rm C}
\overline{\nu_L}^T \quad
{\rm CP} \nu_R ({\rm CP})^{\dagger}=W^\prime
\gamma^0 {\rm C} \overline{\nu_R}^T
\label{cp}
\end{eqnarray}
where $U^\prime$, $V^\prime$, $W^\prime$
are unitary matrices acting in flavour space
and where for notation simplicity we have dropped here the
superscript 0 in the fermion fields.
Invariance of the mass terms under the above CP transformation,
requires that the following relations have to be satisfied:
\begin{eqnarray}
W^{\prime T} M W^\prime &=&-M^*  \label{cpM} \\
U^{\prime \dagger} m W^\prime &=& m^*  \label{cpm} \\
U^{\prime \dagger} m_l V^\prime &=& {m_l}^* \label{cpml}
\end{eqnarray}
From Eqs.~(\ref{cpm}), (\ref{cpM}), one obtains:
\begin{eqnarray}
W^{\prime \dagger}h W^\prime &=& h^* \nonumber  \\
W^{\prime \dagger}H W^\prime &=& H^* \label{wh}
\end{eqnarray}
with $h=m^{\dagger}m$, $H=M^{\dagger}M$.
It can be then readily derived, from Eqs.~(\ref{cpM}) and
(\ref{wh}), that CP invariance requires:
\begin{eqnarray}
I_1 \equiv {\rm Im Tr}[h H M^* h^* M]=0 \nonumber
\label{i1} \\
I_2 \equiv {\rm Im Tr}[h H^2 M^* h^* M]=0  \nonumber
\label {i2l} \\
I_3 \equiv {\rm Im Tr}[h H^2 M^* h^* M H]=0
\label{i3l}
\end{eqnarray}
By construction, these WB invariants are only sensitive to the CP
violating phases which appear in leptogenesis. This is due to the
fact that m always appears in the combination $m^{\dagger}m$.
WB invariant conditions are particularly useful because
they can be evaluated and analysed in any conveniently
chosen WB (see Ref. \cite{Branco:2001pq} for further discussion
of these conditions).
Since there are six independent CP violating phases, one may wonder
whether
one can construct other three independent WB invariants, apart from
$I_i$,
which would describe CP violation in the leptonic sector. This is indeed
possible, a simple choice are the WB invariants ${\bar I}_i (i=1,2,3)$,
obtained
from $I_i$, through the substitution of $h$ by
${\bar h}=m^{\dagger} h_l m$,
where $h_l=m_l {m_l}^{\dagger}$. For example one has:
\begin{equation}
{\bar I_1}={\rm Im Tr}(m^{\dagger}h_l m H M^* m^T {h_l}^* m^* M)
\label{ibar}
\end{equation}
and similarly for $\bar{I_2}, \bar{I_3}$. As it was the case for $I_i$,
CP invariance requires that $\bar{I_i}=0$.

\section{\bf CP violating phases relevant at low energies}
The neutrino mass matrix $\cal M$
is diagonalized by the transformation:
\begin{equation}
V^T {\cal M}^* V = \cal D \label{dgm}
\end{equation}
where ${\cal D} ={\rm diag.} (m_{\nu_1}, m_{\nu_2}, m_{\nu_3},
M_{\nu_1}, M_{\nu_2}, M_{\nu_3})$,
with $m_{\nu_i}$ and $M_{\nu_i}$ denoting the physical
masses of the light and heavy Majorana neutrinos, respectively. It is
convenient to write $V$ and $\cal D$ in the following form:
\begin{eqnarray}
V= \left (\begin{array}{cc}
K & R \\
S & T \end{array}\right) ; \ \ \
{\cal D}=\left(\begin{array}{cc}
d & 0 \\
0 & D \end{array}\right) .
\end{eqnarray}
From Eq. (\ref{dgm}) one obtains to an excellent approximation:
\begin{eqnarray}
S^\dagger=-K^\dagger m M^{-1} \label{13} \\
-K^\dagger m \frac{1}{M} m^T K^* =d \label{14}
\end{eqnarray}
Eq.(\ref{14}) is the usual seesaw formula. In this approximation K
is a unitary matrix corresponding to the three-by-three
Maki-Nakagawa-Sakata matrix \cite{Maki:1962mu}.
The neutrino weak-eigenstates are related to the mass eigenstates by:
\begin{equation}
{\nu^0_i}_L= V_{i \alpha} {\nu_{\alpha}}_L=(K, R)
\left(\begin{array}{c}
{\nu_i}_L  \\
{N_i}_L \end{array} \right) \quad \left(\begin{array}{c} i=1,2,3 \\
\alpha=1,2,...6 \end{array} \right)
\label{15}
\end{equation}
and thus the leptonic charged current interactions are given by:
\begin{equation}
- \frac{g}{\sqrt{2}} \left( \overline{l_{iL}} \gamma_{\mu} K_{ij}
{\nu_j}_L +
\overline{l_{iL}} \gamma_{\mu} R_{ij} {N_j}_L \right) W^{\mu}+h.c.
\label{16}
\end{equation}
From Eqs.(\ref{15}), (\ref{16}) it follows that $K$ and $R$ give the
charged current
couplings of charged leptons to the light
neutrinos $\nu_j$ and to the heavy
neutrinos $N_j$, respectively.  In this approximation,
with $K$ a unitary matrix,
it is clear that we can rotate away three phases
on the left by a redefinition of the physical charged leptonic
fields so that $K$ is left with three physical CP violating
phases, one of Dirac type, $\delta$,  and two of Majorana type,
which have an interesting geometrical interpretation
in terms of unitarity triangles \cite{Aguilar-Saavedra:2000vr}.
These are the three CP violating phases relevant
at low energies. On the other hand
from Eq.(\ref{dgm}), taking into account the zero entry in ${\cal M}$,
one derives the following exact relation:
\begin{equation}
R=m T^* D^{-1} \label{exa}
\end{equation}
In the WB where the right-handed Majorana neutrino mass
is diagonal it follows to an excellent approximation that:
\begin{equation}
R=m D^{-1} \ \ \ \ \  {\rm or}\ \ {\rm else} \ \ \ \
R_{ij} D_j = m_{ij} \label{app}
\end{equation}
leading to:
\begin{equation}
A^j =  \frac{g^2}{{M_W}^2} \sum_{k \ne j} \left[ (M_k)^2
{\rm Im} \left((R^\dagger R)_{jk} (R^\dagger R)_{jk} \right)
\frac{1}{16 \pi} \left(I(x_k)+ \frac{\sqrt{x_k}}{1-x_k} \right)
\right]
\frac{1}{(R^\dagger R)_{jj}}.
\end{equation}
From the first equality in Eq.(\ref{app}) and Eq.(\ref{muh})
we see, once again, that only the phases $\sigma$, $\beta _1$
and $\beta _2$ are relevant for leptogenesis. Furthermore
the first equality in Eq.(\ref{app}) implies that the two CP
violating phases $\beta _1$ and $\beta _2$ in m (see Eq.(\ref{muh}))
appear in Eq.(\ref{16}) as Majorana type phases since $P_\beta $
commutes with $D^{-1}$ and, as a result, these phases can be
shifted to the physical heavy neutrino masses.

\section{\bf Relating CP violation in leptogenesis with CP violation
at low energies}
In this section we address the question of whether it is possible to
infer about the size of CP violation at low energies from the size and
sign of the observed BAU. For definiteness, let us consider the
parametrization of $m$ given before, where the six phases are
$\rho$, $\alpha_1$, $\alpha_2$, $\sigma$, $\beta_{1}$, $\beta_{2}$.
We have already seen that leptogenesis is controlled by the phases
$\sigma$, $\beta_{1}$, and $\beta_{2}$. On the other hand the phases
relevant at low energies are those appearing in $K$ and
resulting from the diagonalization of the effective left-handed
neutrino mass matrix given by:
\begin{equation}
m_{ef}=-m \frac{1}{D} m^T
\label{mef}
\end{equation}
The strength of CP violation at low energies, observable
for example through neutrino
oscillations, can be obtained from the following low-energy
WB invariant:
\begin{equation}
Tr[h_{ef}, h_l]^3=6i \Delta_{21} \Delta_{32} \Delta_{31}
{\rm Im} \{ (h_{ef})_{12}(h_{ef})_{23}(h_{ef})_{31} \} \label{trc}
\end{equation}
where $h_{ef}=m_{ef}{m_{ef}}^{\dagger},\  h_l=m_l {m_l}^{\dagger}$ and
$\Delta_{21}=({m_{\mu}}^2-{m_e}^2)$ with analogous expressions for
$\Delta_{31}$, $\Delta_{32}$. CP violation in neutrino oscillations
\cite{osc} is only affected by the phase $\delta$.
The important point is that the phase
$\delta$ is, in general, a function of all the six phases $\rho$,
$\alpha_1$, $\alpha_2$,
$\sigma$, $\beta_{1}$, $\beta_{2}$ as can be seen
from Eq.~(\ref{14}). Since leptogenesis only depends
on $\sigma $, $\beta_{1}$ and $\beta_{2}$,
it is clear that, in general, one cannot directly relate the size
of CP violation responsible for leptogenesis with the strength
of CP violation at low energies. Yet it can be seen by
computing the invariant  $Tr[h_{ef}, h_l]^3$
that in a model where the leptonic
mass matrices are constrained (e. g. by flavour symmetries) so that
only one of the phases (for example $\sigma$ ) is non-vanishing,
one can establish a direct connection between the size of the observed
BAU and the strength of CP violation at low energies observable,
for example, in neutrino oscillations.
Of special interest are specific GUT inpired scenarios such as
the case of m given by:
\begin{equation}
m=dU_{R}  \label{dur}
\end{equation}
where d is diagonal and $U_R$ is a generic unitary matrix, in a
WB where $m_l$ and M are both real and diagonal. This case has
been discussed with all generality in Branco et al. in
Ref. \cite{link}.
Another interesting possibility are models with spontaneous CP
violation at a high energy scale such as the one
presented in Ref. \cite{Branco:2001pq}.

\section{\bf Concluding remarks}
This talk is based on a more detailed work \cite{Branco:2001pq}
where we studied the possible sources of CP violation in
the minimal seesaw model and addressed the question of
whether it is possible to establish a connection between
CP violation responsible for leptogenesis and CP violation
observable at low energies. It was shown that, in general,
such a connection does not exist but there are special
interesting scenarios where it may be established.

\section*{\bf Acknowledgments}
Contributed paper based on Ref. \cite{Branco:2001pq}
The authors of Ref. \cite{Branco:2001pq}
thank the CERN Theory Division for hospitality during its
preparation and also T. Endoh, R. Gonz\' alez Felipe,
T. Onogi and A. Purwanto for useful discussions.
The work of TM was supported by a fellowship for
Japanese Scholar and Researcher abroad from the
Ministry of Education, Science and Culture of Japan.
The work of BMN was supported by Funda\c c\~ ao para a
Ci\^ encia e a Tecnologia (FCT)
(Portugal) through fellowship SFRH/BD/995/2000; GCB, BMN and MNR
received partial support during the preparation of
Ref. \cite{Branco:2001pq} from FCT through
Project CERN/P/FIS/40134/2000,
Project POCTI/36288/FIS/2000 and Project CERN/C/FIS/40139/2000.
The participation of MNR at the RTN Meeting ``Across the Present
Energy Frontier: Probing the Origin of Mass'', Corfu2001,
where this talk was presented, was partially supported by
the EC under contract HPRN-CT-2000-00148
and by the Project CERN/P/ FIS/40134/2000.
Finally we specially thank the Organizers of Corfu2001
for the stimulating Meeting.

\end{document}